\documentclass[12pt,epsfig]{article}
\usepackage{bm}
\usepackage{graphicx}
\usepackage{wrapfig}
\usepackage{amsmath}
\usepackage{wick}

\begin{document}

\begin{center}
{\Large\bf Hadron-quark matter phase transition in neutron stars}

\vspace{1cm}

%{Tomoki ~Endo}
{Tomoki~Endo\footnote{*endo@ruby.scphys.kyoto-u.ac.jp}$^{,*}$, 
Toshiki~Maruyama$^2$, Satoshi~Chiba$^2$ and Toshitaka~Tatsumi$^1$}
\vspace{0.5cm}

$^1$ Department of Physics, Kyoto University, Kyoto 606-8502, Japan \\
$^2$ Advanced Science Research Center, Japan
        Atomic Energy Research Institute, Tokai, Ibaraki 319-1195, Japan \\

\end{center}

\vspace{1cm}

\begin{abstract}
A structured mixed phase consisting of quark and hadron phases is
 numerically studied with
the Coulomb screening effect and the surface effect. We carefully introduced the
Coulomb potential, so that a geometrical structure becomes mechanically unstable
when the surface tension is large. Charge densities are largely
 rearranged by the screening effect, and thereby the equation of state
 shows the similar behavior to that given by the Maxwell construction. Therefore,
  although bulk calculations with the Gibbs conditions show that the mixed phase may exist in
 a wide density region, we can see it is restricted to a
 narrow density region by the surface effect and the Coulomb screening
 effect. 
%We can see that Maxwell construction, which seems to be not valid at a first glance, is effectively useful.  
\end{abstract}

\section{Introduction} \indent

 It has been believed that hadron matter changes to quark matter at
 high-density  region by way of the ``deconfinement phase
 transition''.  Unfortunately the deconfinement phase transition have not 
been well 
 understood up to now, and many authors have studied it 
 by model calculations or by first-principle calculations like lattice QCD.
 These studies are now developing, and many exciting results have been 
reported.
 Properties of quark matter have been 
 actively studied theoretically in quark-gluon
 plasma, color superconductivity \cite{alf1,alf3} or magnetism \cite{tat1,tat2,tat3}, and experimentally in relativistic
 heavy-ion collision (RHIC), HERA or early universe and compact stars
 \cite{mad3,chen}. 

\begin{figure}[htb]
\begin{minipage}[t]{80mm}
%\includegraphics[width=8cm]{edens_rho.eps}
%\begin{center}
\includegraphics[width=80mm]{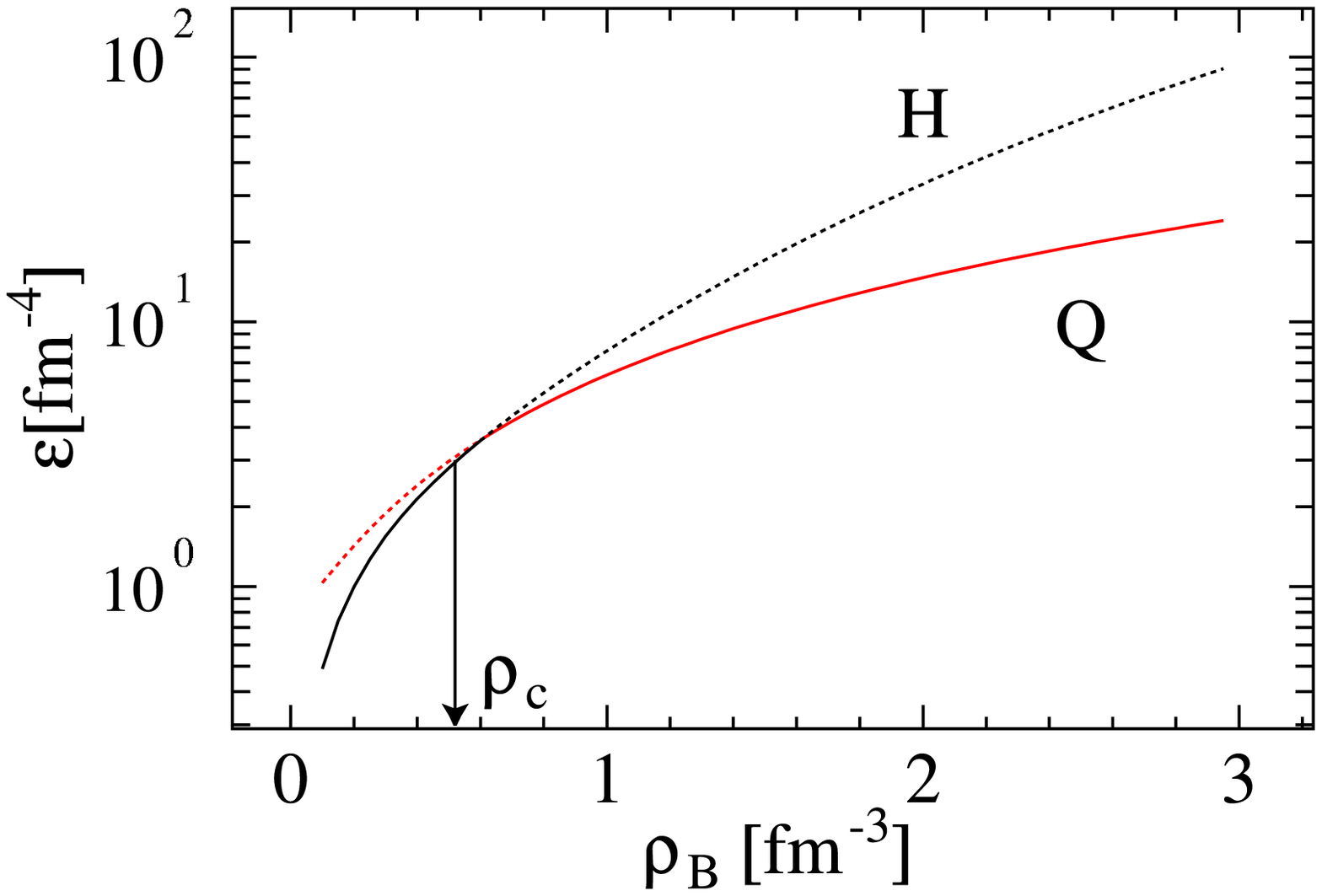}
\caption{Energy density $\epsilon$ for uniform hadron matter and quark
		   matter. Uniform quark matter is energetically
		   favorable in high-density region, while hadron matter
		   in low-density region.}
\label{eosbulk}
\end{minipage}
\hspace{8pt}
\begin{minipage}[t]{80mm}
\includegraphics[width=80mm]{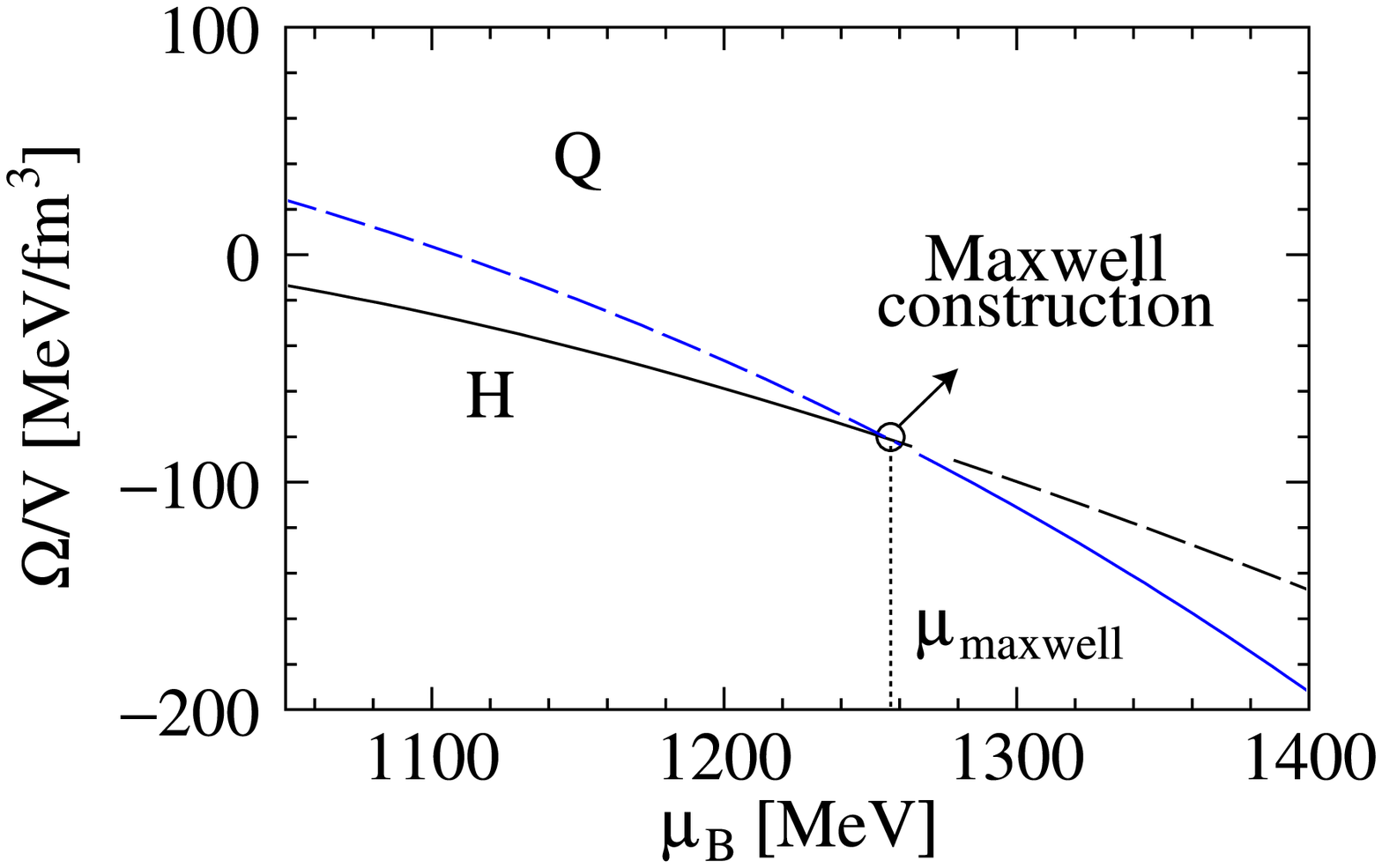}
\caption{ Thermodynamic potential $\Omega$ of uniform hadron matter and quark
		   matter as a function of baryon chemical potential $\mu_\mathrm{B}$. Uniform quark matter is energetically
		   favorable in high $\mu_\mathrm{B}$ region while hadron matter
		   in low $\mu_\mathrm{B}$ region.}
\label{omebulk}
\end{minipage}
%\end{center}
\end{figure}

 When we calculate uniform hadron matter (nuclear matter) and quark
 matter at zero temperature separately, by using the MIT bag model, we can expect the
 first-order phase transition as seen in Figs.\ \ref{eosbulk} and \ref{omebulk}.
 We can see that quark matter is an energetically favorable
 state at high-density region, $\rho > \rho_\mathrm{c}$ (Fig.\ \ref{eosbulk}). As we can see in Fig.\ \ref{omebulk}, the
 thermodynamic potential of quark matter becomes lower at higher 
  baryon-number chemical potential.  These results suggest the deconfinement
 phase transition at high densities.

 The features of the deconfinement phase transition
 have not been fully
 elucidated yet. We assume here that it is the first order phase transition,
and use the bag model for simplicity. 
 Then the thermodynamically forbidden region appears in the equation 
of state (EOS) and we can expect the mixed phase, 
 the hadron-quark mixed phase, in some density region, which may exist in inner core region of 
neutron stars 
 and during the hadronization of high-temperature quark-gluon plasma at RHIC 
experiment. 
 We have to apply 
 the Gibbs conditions (GC) to get EOS in thermodynamic
 equilibrium: GC demand chemical equilibrium, pressure valance and
 thermal equilibrium between two phases;
\begin{equation}
 \mu_{\mathrm{B}}^{\mathrm{quark}} = \mu_{\mathrm{B}}^{\mathrm{hadron}} (\equiv \mu_{\mathrm{B}} ),
 \hspace{5pt} \mu_{\mathrm{Q}}^{\mathrm{quark}} =
 \mu_{\mathrm{Q}}^{\mathrm{hadron}}  (\equiv \mu_\mathrm{e} ), \hspace{5pt} P^{\mathrm{quark}}=P^{\mathrm{hadron}}, \hspace{5pt} T^{\mathrm{quark}} = T^{\mathrm{hadron}},
\label{gc}
\end{equation}
where $\mu_{\mathrm{B}}^i$ and $\mu_{\mathrm{Q}}^i$ are baryon-number
and charge chemical potentials, respectively. Thermal equilibrium is
implicitly achieved at $T=0$.
Note that there are two independent chemical potentials in this phase transition. 
%It could be seen that this system should be much different from the liquid-vapor 
In such a case the system should be much different from the liquid-vapor 
phase transition which is described by only  one chemical potential. 

%\begin{wrapfigure}{r}{60mm}
%  \epsfxsize = \halftext
%  \centerline{ \epsfbox{edens_rho.eps}}
%\end{wrapfigure}

 These GC must be fulfilled in the hadron-quark mixed phase. 
 On the other hand,
 the Maxwell construction (MC) may be very familiar and has been used by many
 authors to get EOS for the first order phase transitions \cite{weis,migd,elli,rose}.
 It is well known that MC is a correct prescription to derive EOS for the 
liquid-vapor phase transition. 
% MC treat uniform matters and charge neutrality is locally established.
% Thus we call it ``local charge neutrality'' \cite{gle3}.
% %If we use this MC to hadron-quark mixed phase.
%
 However, Glendenning \cite{gle1} pointed out that MC is not appropriate
 for the
 hadron-quark mixed phase: one of GC about the charge chemical 
equilibrium,
 $ \mu_{\mathrm{Q}}^{\mathrm{quark}} = \mu_{\mathrm{Q}}^{\mathrm{hadron}}$, 
 is not satisfied in MC, since the local charge neutrality is implicitly assumed  without imposing this condition. He emphasized that 
 the local charge neutrality is too restrictive and 
each hadron or quark phase may have a net charge 
 because only the total charge must be kept neutral.

%\begin{wrapfigure}{r}{60mm}
%  \epsfxsize = \halftext
%  \centerline{ \epsfbox{edens_rho.eps}}
%\begin{figure}[h!]\begin{center}
%\includegraphics[width=8cm]{edens_rho.eps}
%\begin{center}
%\includegraphics[width=60mm]{uniform.eps}
%\caption{ Thermodynamic potential of uniform hadron matter and quark
%		   matter. Uniform quark matter is energetically
%		   favorable in high value of baryon chemical potential where hadron matter
%		   is favored in low.}
%\label{fig1}
%\end{center}
%\end{figure}
%\end{wrapfigure}

 When we use GC in the bulk calculation, which we explain in
 detail later, 
 we can see the mixed phase appears in a large density region. 
 %there is no constant pressure area, 
 The pressure is not constant as density changes in the mixed phase,
 while only a constant pressure is obtained from MC as shown in
 Refs.\ \cite{gle1,gle3}.
 
However, the bulk calculation is too simple and Heiselberg 
et al.\ \cite{peth} claimed the importance of including the finite-size 
effects, i.e., 
 the surface tension at the hadron-quark boundary and the Coulomb interaction 
energy. 
 They studied the quark droplet immersed in hadron matter and found that it is 
 energetically unfavorable if the surface tension is large enough. 
 On the contrary, however, if the surface tension is not large 
 the mixed phase can exist in some density region.
 %in their study.

  Glendenning and Pei \cite{gle2} suggested the crystalline structure by a
 bulk calculation using the small surface tension:  
 one phase is immersed in another phase with
 various geometrical structures; ``droplet'', ``rod'', ``slab'', ``tube'', and ``bubble''. 
 These are called the structured mixed phases (SMP). 
 Applying the results based on the bulk calculation to neutron stars,
 they suggested that there could develop SMP in
 the core region for several kilo meters in 
thickness.

\begin{wrapfigure}{r}{120mm}
%  \epsfxsize = \halftext
%  \centerline{ \epsfbox{edens_rho.eps}}
%\begin{figure}[h!]\begin{center}
%\includegraphics[width=8cm]{edens_rho.eps}
%\begin{center}
\includegraphics[width=120mm]{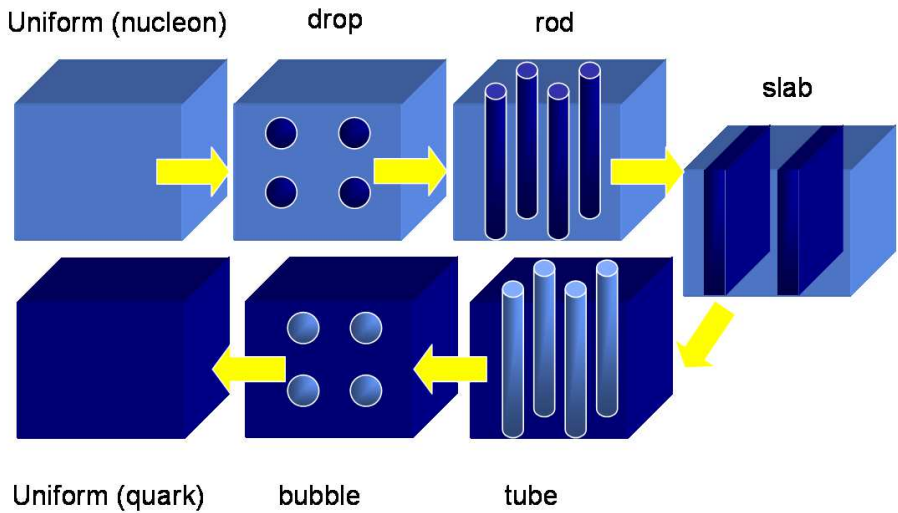}
\caption{ Schematic view of the structured mixed phases.}
\label{smp}
%\end{center}
%\end{figure}
\end{wrapfigure}
At a first glance, this view seems to be reasonable and there 
may appear the mixed phase in
a large density region. However, there are still many points to be elucidated
 about the finite-size effects. 
Voskresensky et al.\ emphasized that the proper treatment of the 
Coulomb interaction is important in the mixed phase  \cite{vosk}.

%
%  On the other hand, there is a question that the picture derived from MC
% is really meaningless or not. MC is well known as valid method in liquid-vapor
% phase transition. We have already said that MC in this hadron-quark phase
% transition is not appropriate, but we wonder if the picture derived from
% MC is really meaningless or have some physical meaning.
%
%

% First, we have to carefully consider the
%Coulomb interaction to treat the mixed phase of charged
%matter. 
First note that there is the relation between chemical potential and the 
Coulomb potential by way of the gauge transformation: chemical potential is 
not well defined before the gauge fixing.  
Secondly, the charge chemical equilibrium could be rather 
satisfied even in MC, once the Coulomb potential $V_\mathrm{Coul}$ is 
incorporated: 
\begin{equation}
\hfill \mu_{\mathrm{Q}}^{\mathrm{quark}}\ = \mu_{\mathrm{Q}}^{\mathrm{hadron}}, %( \equiv \mu_{\mathrm{e}} )
\label{mc1}
\end{equation}
whereas 
\begin{equation}
\mu_{\mathrm{Q}}^{\mathrm{quark}} - V_{\mathrm{Coul}}^{\mathrm{quark}} \neq
 \mu_{\mathrm{Q}}^{\mathrm{hadron}} - V_{\mathrm{Coul}}^{\mathrm{hadron}}.
\label{mc2}
\end{equation}
Note that the electron chemical potential $\mu_e=\mu_Q$ and 
the electron number density can be expressed in terms of the combination,  
$\left( \mu_\mathrm{Q} -V_\mathrm{Coul}\right)$ (see Eq.\ (48)), when the 
Coulomb potential $V_{\rm Coul}$ is introduced.
%,  corresponds to
%electron number density \cite{vosk}. 
Therefore, Eqs.\ (\ref{mc1}), (\ref{mc2}) mean that the charge chemical potential is
equal, while the electron number is different between the hadron and quark 
phases. Equation (\ref{mc2}) is reduced to 
$\mu_e^{\rm quark}\neq \mu_e^{\rm hadron}$ in the absence of the Coulomb 
potential, which is the previous claim that the charge chemical 
equilibrium is apparently violated in MC.
Thus we can see that the previous claim means nothing but the difference 
of the electron number in two phases \cite{vosk}.
Thirdly,  it is important
to take into account the
Coulomb screening effect\footnote{The Coulomb
 screening effect on kaon condensation is studied by Norsen and Reddy
 \cite{nors}.}. We have to solve the Poisson equation
consistently with other equations of motion for charged particles.
Actually, Voskresensky et al.\  showed SMP is mechanically unstable by the Coulomb
screening effect when the surface tension is not
small. If SMP is mechanically unstable, 
the phase transition should be similar to that in MC.
% Voskresensky et al.\ \cite{vosk} studied the Coulomb screening effect in
% the hadron-quark mixed phase 
% and pointed out 
% the importance to solve the Poisson equation consistently.
% It has been found
% that SMP becomes mechanically unstable by the
% Coulomb screening effect if the surface tension is not
% small \cite{vosk}, 

% In their study, however, 
 %Voskresensky et al.\ 
% a linear approximation (RPA) was employed to solve the Poisson equation analytically.
% As the Coulomb potential and charge densities are the
% functions of each other, it is difficult to solve it analytically
% without any approximation. 
 %in their formalism.  
 It can be easily seen that there should occur
 rearrangement of charged particle densities by the Coulomb
 interaction. On the contrary such rearrangement modifies the Coulomb potential.
 As a result the Poisson equation becomes highly non-linear and difficult to solve analytically. 
 In their study, 
 a linear approximation (RPA) was employed to solve the Poisson equation analytically.
 It is valid to use the approximation for pointing out the important property of SMP, 
 but it may be conceivable that various charge properties like ``global charge neutrality'' 
 for GC, ``local charge neutrality'' for MC, and ``the Coulomb screening effect'' 
 have important roles in the mixed phase. 
 Therefore, solving the Poisson equation
 without any approximation is of much significance. 
 We have reported a preliminary result for the case of droplet
 \cite{end1}. It would be also very interesting to 
derive EOS in our consistent calculation for the hadron-quark matter phase transition.  

\section{Bulk calculation with the Gibbs conditions} \indent

 Now we consider
 two infinite matters separated by a sharp
 boundary: uniform quark matter and hadron matter.
 We consider the quark phase consists of u, d, s quarks and electron,
 and the
 hadron phase proton, neutron and electron. We discard the Coulomb interaction 
in the calculation.  Then we can evaluate the total 
thermodynamic
 potential $\Omega_\mathrm{tot}$:
\begin{equation}
\Omega_{\mathrm{tot}} = \Omega_{\mathrm{u}} + \Omega_{\mathrm{d}} + \Omega_{\mathrm{s}} + \Omega_{\mathrm{n}} + \Omega_{\mathrm{p}} +\Omega_\mathrm{e}. 
%E_{\mathrm{V_{\mathrm{Coul}}}} &=& \frac{1}{2} \int d^3r d^3r^{\prime} \frac{Q_i \rho_i (\bm{r}) Q_j \rho_j (\bm{r}^{\prime} )}{\left| \bm{r} - \bm{r}^{\prime} \right|}. 
\end{equation}
The explicit expressions of $\Omega_i$ for non-uniform matter are given in the next section. We use them by 
replacing the space dependent quantities by constants for each uniform matter. 

 Introducing the volume fraction of the quark
 phase $f$, we impose the global charge neutrality:
%in which each phase can have net 
% charge. 
 the total charge density vanishes,
\begin{equation}
 f \rho_{\mathrm{Q}} + (1-f) \rho_{\mathrm{H}} = 0,
\label{totchcond}
\end{equation}
where $ \rho_{\mathrm{Q}}$ and $ \rho_{\mathrm{H}}$ are
the net charge densities of the quark and hadron phases, respectively, 
%These are defined as
\begin{equation}
\begin{split}
\rho_{\mathrm{Q}}&=\frac{2}{3}\rho_\mathrm{u} - \frac{1}{3}\rho_\mathrm{d} -
 \frac{1}{3}\rho_\mathrm{s} - \rho_\mathrm{e}, \\
\rho_{\mathrm{H}}&=\rho_\mathrm{p}-\rho_\mathrm{e}. 
\end{split}
\end{equation}

 Note that we
 have now six chemical potentials; $\mu_\mathrm{u}$, $\mu_\mathrm{d}$,
 $\mu_\mathrm{s}$, $\mu_\mathrm{p}$, $\mu_\mathrm{n} (\equiv
 \mu_\mathrm{B})$, $\mu_\mathrm{e}$. We first consider $\beta$ equilibrium
 in each phase and chemical
 equilibrium at the hadron-quark boundary. Thus all chemical potentials are constant. 
The chemical equilibrium conditions then are:
\begin{eqnarray}
 \mu_\mathrm{u}+\mu_\mathrm{e} = \mu_\mathrm{d}, \label{chemequ}\\
 \mu_\mathrm{d}=\mu_\mathrm{s} , \label{chemeqs}
\end{eqnarray}
in the quark phase,
\begin{equation}
 \mu_\mathrm{p}+\mu_\mathrm{e}= \mu_\mathrm{n} \label{chemeqb}
\end{equation}
in the hadron phase, and
\begin{eqnarray}
 \mu_\mathrm{n} = \mu_\mathrm{u}+2 \mu_\mathrm{d}, \label{chemeqn}\\
 \mu_\mathrm{p} = 2 \mu_\mathrm{u}+\mu_\mathrm{d}, \label{chemeqp}
\end{eqnarray}
at the hadron-quark interface.
 The last condition (\ref{chemeqp}) can be derived from other four conditions,
 so that there are left four independent conditions for chemical equilibrium.
Therefore, if we give two chemical potentials $\mu_\mathrm{B}$ and
$\mu_\mathrm{e}$ by hand, 
%as for there are four unknown values of the chemical potentials
%and four equations, 
we can determine these four chemical potentials; $\mu_\mathrm{u}$, 
$\mu_\mathrm{d}$, $\mu_\mathrm{s}$ and $\mu_\mathrm{p}$ from these
 four equations.

Next, we can determine $\mu_\mathrm{e}$ by the global charge neutrality
condition (\ref{totchcond}). $f$ is still unknown at this point
and finally, we find the optimal value of $f$ by using one of GC;
$P^{\mathrm{quark}}=P^{\mathrm{hadron}}$, where pressure $P$ 
is given by the thermodynamic relation: $ P=-\Omega/V$.
Thus once
$\mu_\mathrm{B}$ is given, all other values $\mu_i$ ($i=u,d,s,p,e$)
and $f$ can be obtained.
 
\begin{wrapfigure}{r}{80mm}
%  \epsfxsize = \halftext
%  \centerline{ \epsfbox{edens_rho.eps}}
%\begin{figure}[h!]\begin{center}
%\includegraphics[width=8cm]{edens_rho.eps}
%\begin{center}
\includegraphics[width=80mm]{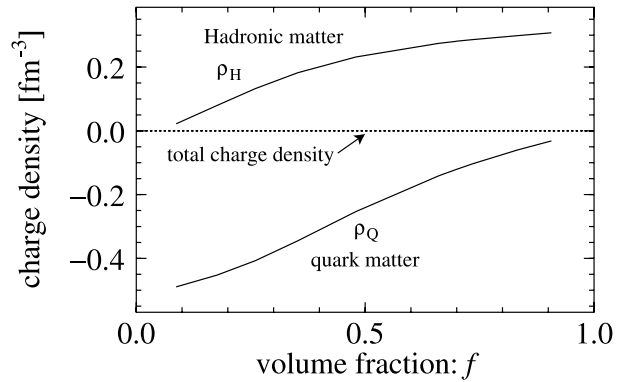}
\caption{ Charge density of each phase. 
The total charge density is always zero.}
\label{chdens}
%\end{center}
%\end{figure}
\end{wrapfigure}

 We show the results of the bulk calculation.
We present the charge densities in Fig.\ \ref{chdens}. We can see the total charge density is
zero, while the quark phase is negatively charged and the hadron phase
positively charged. Note that MC always gives a null charge density in each 
phase due to the local charge neutrality \cite{gle3}.
%if we draw the results of MC in Fig.\
%\ref{chdens}, charge density of each phase is always zero because MC treat
%uniform matters \cite{gle3}.

\begin{figure}[htb]
\begin{minipage}[t]{80mm}
%
%\begin{wrapfigure}{r}{80mm}
%  \epsfxsize = \halftext
%  \centerline{ \epsfbox{edens_rho.eps}}
%\begin{figure}[h!]\begin{center}
%\includegraphics[width=8cm]{edens_rho.eps}
%\begin{center}
\includegraphics[width=80mm]{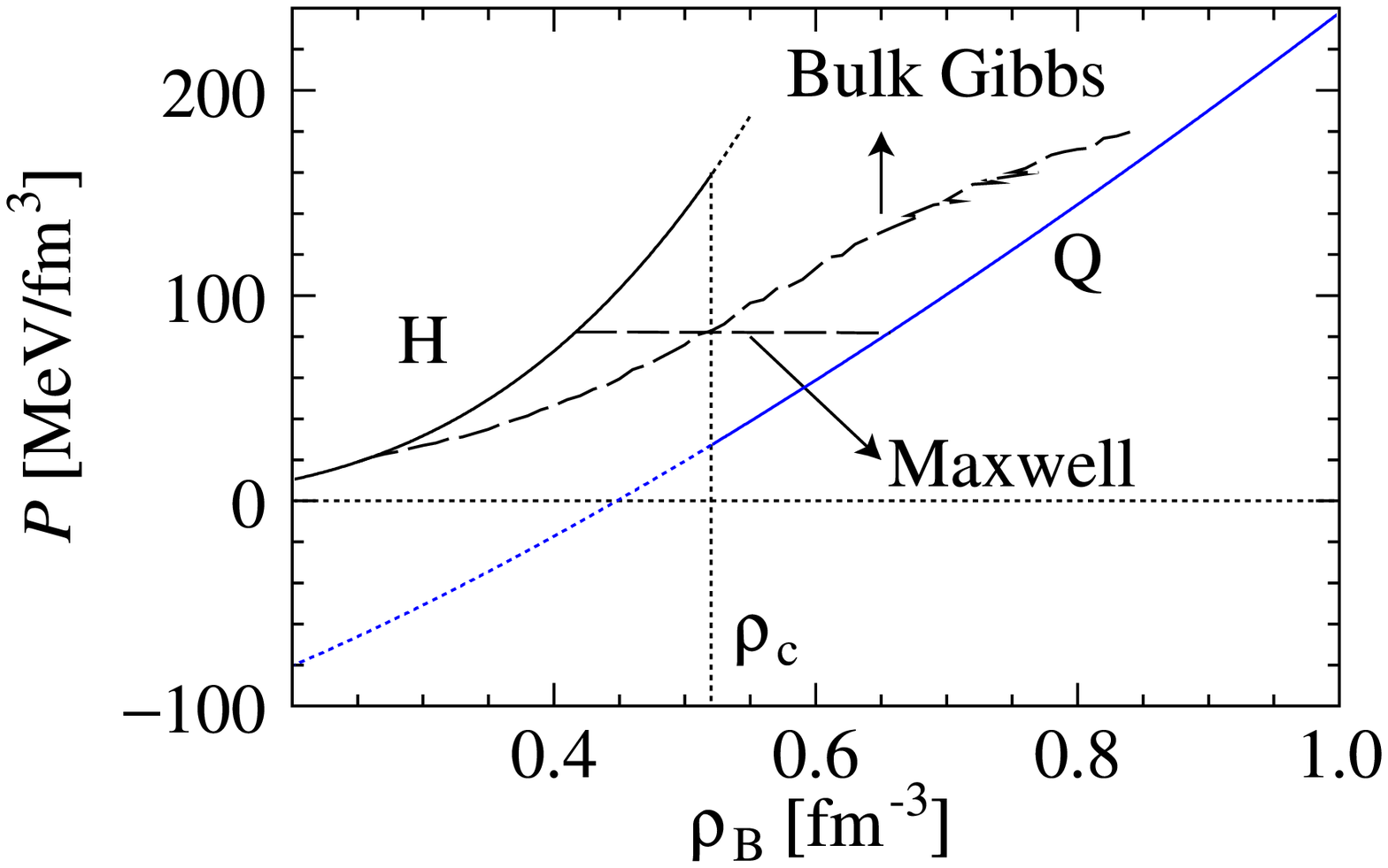}
\caption{ Pressure of uniform matter, that given by the bulk calculation with 
GC (Bulk Gibbs) and that given 
 by MC (Maxwell).}
\label{presbulk}
\end{minipage}
\hspace{8pt}
\begin{minipage}[t]{80mm}
%\end{center}
 %\end{figure}
%\end{wrapfigure}
%

%\begin{wrapfigure}{r}{80mm}
%  \epsfxsize = \halftext
%  \centerline{ \epsfbox{edens_rho.eps}}
%\begin{figure}[h!]\begin{center}
%\includegraphics[width=8cm]{edens_rho.eps}
%\begin{center}
\includegraphics[width=80mm]{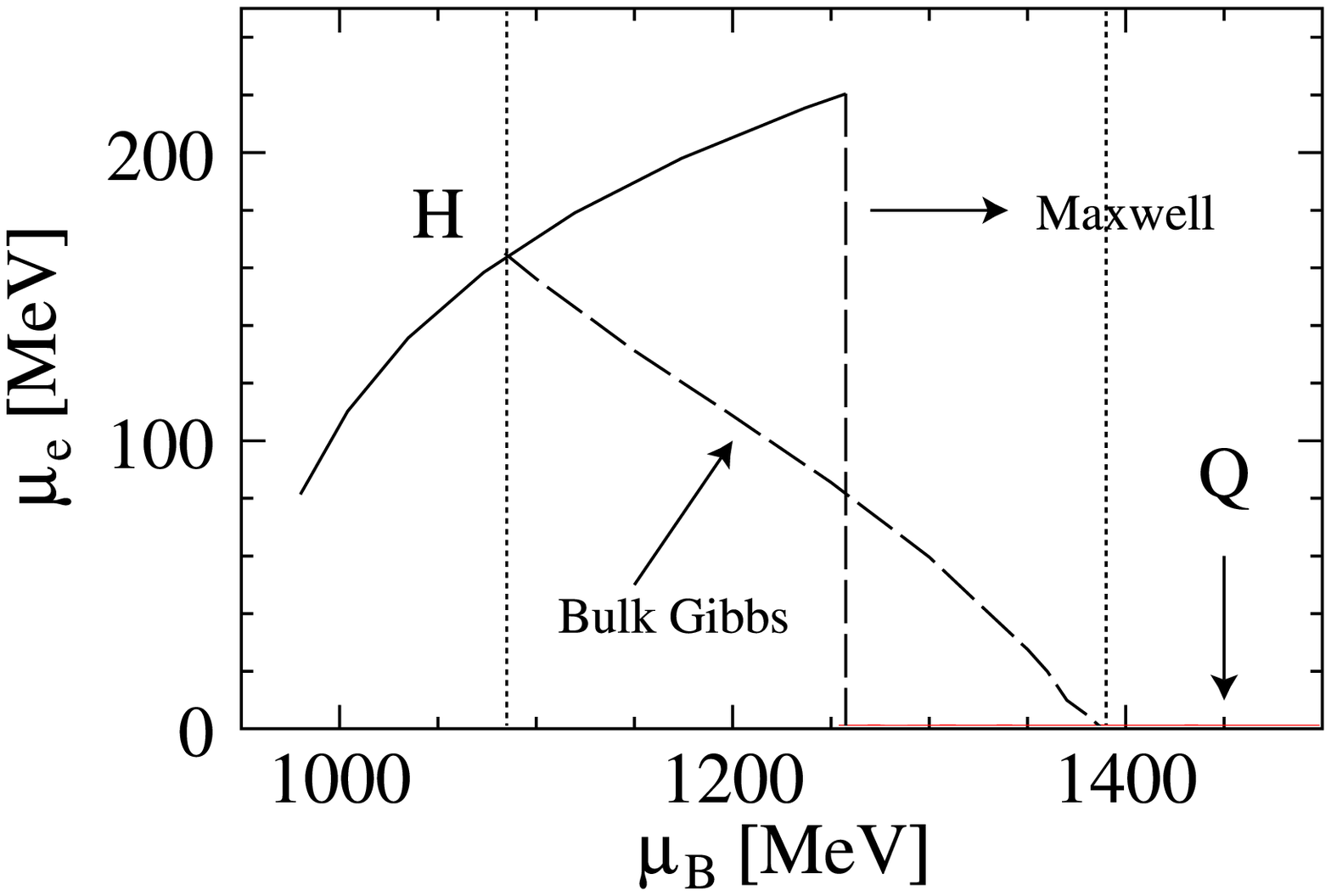}
\caption{ Phase diagram in the chemical potential plane. H means uniform hadron
 matter and Q uniform quark matter. Other notations have the same
 meaning as in Fig.\ \ref{presbulk}}
\label{mubmuebulk}
%\end{center}
\end{minipage}
\end{figure}

Figure \ref{presbulk} shows pressure versus baryon-number density. 
The most important difference between ``Maxwell'' and ``Bulk Gibbs'' is that the pressure given by 
MC is
constant while that given by the bulk calculation with GC is density dependent.
 Remember that the outstanding feature in these two results comes 
from that each
 hadron or quark phase has a net charge in ``Bulk Gibbs'', while 
 each phase is neutral in ``Maxwell''. 
%This difference change the picture of hadron-quark mixed phase. 

The phase diagram in the $\mu_{\rm B}-\mu_{\rm e}$ plane is presented 
in Fig.\ \ref{mubmuebulk}.
We can easily see that the charge chemical potential $\mu_{\mathrm{e}}$ is much different between
uniform hadron matter and quark matter for MC, which means 
%  We can easily see that one of the Gibbs
% condition; 
chemical equilibrium condition for $\mu_{\mathrm{e}}$ is not
 satisfied in MC. Thus it could be said that MC is not a proper way 
in the bulk calculation.
  Glendenning \cite{gle1} pointed out this defect in MC and showed
 another result (Bulk Gibbs) within the bulk calculation. 
 One can easily see that the
 mixed phase appears in a wide $\mu_\mathrm{B}$
 region in Fig.\ \ref{mubmuebulk}. 

%\end{wrapfigure}

After the claim by Glendenning, Heiselberg et al.\ \cite{peth} demonstrated 
that the finite-size effects may disfavor the mixed phase 
%by extending the quark droplet
%embedded in hadron matter with 
by extending the bulk calculation to include the Coulomb interaction and the 
surface energy.
% They showed the droplet is energetically
%favorable state for small surface tension, and pointed out that if
%the surface tension is large enough, the droplet is energetically
%unfavorable.
On the other hand, Glendenning and Pei \cite{gle2} suggested 
``crystalline structures of the mixed phase'' 
%depends
%on this droplet results 
because SMP are energetically more favorable than uniform matter
even including
the surface and Coulomb effects. 
%As these hadron-quark systems corresponds to
%neutron star matter, he suggested that a star have the mixed phase of several kilo meters
%in its radius because there is mixed phase within large
%density region.  
%

Voskresensky et al.\ pointed out the importance of the Coulomb screening
effect and suggested the usefulness of MC for the mixed phase.
We have seen MC is apparently incorrect because if we apply MC to
this mixed phase, one of GC, charge chemical
equilibrium, is violated. However, once the Coulomb interaction is taken into account, there is still a room to satisfy the condition without spoiling the picture given by MC, as noted in the previous section. Since the Coulomb
screening effect had been taken into account 
%and the property of its
%effect was discussed 
with an approximation in their study, we may still
address a question, ``is
the picture derived from MC is completely meaningless when we include
the Coulomb interaction in a proper way?''.
%We have discarded the finite size effects of surface tension and the
%Coulomb energy in the simple bulk calculation, and  seen the mixed
%phase can exist even with small finite size effects. 
All we have to do next is to solve the Poisson equation without any
approximation, strictly satisfying GC.

\section{Formalism}

\subsection{Density Functional Theory} \indent

We use the idea of the Density Functional Theory (DFT) \cite{parr,drez} to
study the quark-hadron mixed phase.
We employ here the Thomas-Fermi approximation, which gives the simple expressions for the energies.
 We summarize below some energy
expressions with the local density approximation (LDA); 
% LDA is the method that we use
the expressions are first derived for uniform system, and then we regard 
densities as the space-dependent functions.
%For the space-dependent quantities 
%we use the expressions of LDA which are derived for uniform system.

First, the kinetic energy is simply expressed as
\begin{eqnarray}
\epsilon (\rho(\bm{r})) &=& \frac{2}{\left(2\pi\right)^3} \int_0^{p_\mathrm{F}(\bm{r})} d^3p \sqrt{p^2+m^2}
\label{kinene} \\
 &=&  \frac{m^4}{8\pi^2} \left[ \frac{p_\mathrm{F}(\bm{r})}{m} \sqrt{1 + \left(\frac{p_\mathrm{F}(\bm{r})}{m} \right)^2} \left( 2 \left( \frac{p_\mathrm{F}(\bm{r})}{m} \right)^2  + 1 \right) \right.\nonumber\\
 && \left.\hspace*{2cm}- \ln \left( \frac{p_\mathrm{F}(\bm{r})}{m} + \sqrt{ 1 + \left( \frac{p_\mathrm{F}(\bm{r})}{m} \right)^2  }   \right)    \right],  \nonumber
\end{eqnarray}
where
$m$ is the particle mass and 
the Fermi momentum $p_\mathrm{F}(\bm{r})$ is expressed by the density
profile $\rho(\bm{r})$ as $p_\mathrm{F}(\bm{r}) = \left( \pi^2 \rho(\bm{r})  \right)^\frac{1}{3}$.

Next, we consider the interaction energy of quarks. 
%When we treat quarks, we have to consider the color degrees of freedom.
 We take the first-order
contribution which comes from the one-gluon exchange interaction in uniform quark matter,
\begin{equation}
\begin{tabular}{c}\includegraphics[width=3cm,height=3cm,keepaspectratio]{oge.eps2}\end{tabular} = \overline{\psi}(p_2) \left( -ig\gamma_\mu \frac{\lambda_a}{2} \right) \psi(p_1) D^{\mu\nu} (p^\prime) \overline{\psi}(p_4)  \left( -ig\gamma_\nu \frac{\lambda_a}{2} \right) \psi(p_3).
\end{equation}
Here $\lambda_a$ is the SU(3) Gell-Mann matrix and $D^{\mu\nu}$ the gluon propagator \cite{tama}.
By way of the Wick contraction, the Hartree and Fock terms are derived:
\begin{equation}
 \wick{22}{\overline{\psi}<1(q) \left( -ig\gamma_\mu \frac{\lambda_a}{2} \right) >1\psi(q) D^{\mu\nu} (0) \overline{\psi}<+(k)  \left( -ig\gamma_\nu \frac{\lambda_a}{2} \right) >+\psi(k)} =\begin{tabular}{c}\includegraphics[width=3cm,height=3cm,keepaspectratio]{hartree.eps2}\end{tabular} 
\end{equation}
for the Hartree term, and
\begin{equation}
 \wick{32}{\overline{\psi}<1(q) \left( -ig\gamma_\mu \frac{\lambda_a}{2} \right) <+\psi(k) D^{\mu\nu} (q-k) \overline{\psi}>+(k)  \left( -ig\gamma_\nu \frac{\lambda_a}{2} \right) >1\psi(q)} = \begin{tabular}{c}\includegraphics[width=3cm,height=3cm,keepaspectratio]{fock.eps2}\end{tabular} 
\end{equation}
for the Fock term.
Due to the traceless property of $\lambda_a$, ${\rm Tr}\lambda_a=0$, the Hartree term gives
a null contribution in the color-singlet quark matter,
\begin{equation}
\wick{2}{\overline{\psi}<+(k)  \left( -ig\gamma_\nu \frac{\lambda_a}{2}	\right)	>+\psi(k)} = \mathrm{Tr} \left[ \overline{\psi}\psi \left( -ig\gamma_\nu \frac{\lambda_a}{2} \right) \right] = 0. 
\end{equation}
Therefore only the Fock term contributes to the interaction energy, which becomes
\begin{eqnarray}
\begin{tabular}{c}\includegraphics[width=2.5cm,height=2.5cm,keepaspectratio]{fock2.eps2}\end{tabular} = -\frac{1}{8}g^2\mathrm{Tr}\left(\lambda_a \lambda_a\right) \int\!\!\!\int \frac{d^4k}{\left(2\pi\right)^4} \frac{d^4q}{\left(2\pi\right)^4} \mathrm{Tr} \left( \gamma_\mu G_D (q) \gamma_\nu G_D (k) \right) D^{\mu\nu}(q-k) \\
= -64 \alpha_s \pi \int\!\!\!\int \frac{d^3q}{(2\pi)^3} \frac{d^3k}{(2\pi)^3} n_q n_k \frac{1}{4k^0q^0} \left( 2m_f^2 -kq \right) \frac{-1}{\left(k-q\right)^2} \equiv \epsilon_{\mathrm{Fock}}.
\end{eqnarray}
Here $G_D$ is the density-dependent quark propagator \cite{tama}. The 
subscript $f$ 
denotes the flavor, $m_f$ the quark mass, $k^0=\left( |\bm{k}|^2 +m_f^2
\right)^{1/2}$ and $n_k=\theta (k_{\mathrm{F}f}-|\bm{k}|) $ with the Fermi
 momentum $k_{\mathrm{F}f}$. 
%To
% incorporate it
%in the density functional, we simply replace the uniform density by the 
%space-dependent one,
%
After some manipulation we easily find
\begin{equation}
\epsilon_{\mathrm{Fock}}  = -\frac{\alpha_s}{\pi^3} \sum_{\mathrm{f}} m_f^4 \left\{ x_{{f}}^4 - \frac{3}{2} \left[ x_{{f}} \eta_{{f}} -\ln\left( x_{{f}}+\eta_{{f}}  \right)  \right]^2  \right\},
\label{fockene}
\end{equation}
where $\displaystyle x_f=\frac{p_{\mathrm{F}f}(\bm{r})}{m_f}$ and
$\eta_f=\sqrt{1+x_f^2}$, and we use the relation $p_\mathrm{Ff}
(\bm{r}) = \left( \pi^2 \rho_f(\bm{r})  \right)^\frac{1}{3}$. 
%###DON'T USE ABBR. (Eq. AND Fig. ETC) AT THE TOP OF A SENTENCE.
Equation (\ref{fockene}) becomes 
$\epsilon_{\rm Fock}=\frac{1}{2} \frac{\alpha_s}{\pi^3} \sum_f p_{Ff}^4$ for 
massless particles.

For the interaction energy of nucleons, we use for simplicity the effective potential
parametrized by densities \cite{vosk},
\begin{eqnarray}
\epsilon_\mathrm{pot} (\bm{r}) &=& S_0 \frac{\left( \rho_\mathrm{n}(\bm{r}) - \rho_\mathrm{p}(\bm{r})  \right)}{\rho_0(\bm{r})} + \left( \rho_\mathrm{n}(\bm{r}) + \rho_\mathrm{p}(\bm{r})  \right) \epsilon_\mathrm{bind} \nonumber \\
&& + K_0 \frac{\left( \rho_\mathrm{n}(\bm{r}) + \rho_\mathrm{p}(\bm{r})  \right)}{18} \left( \frac{\rho_\mathrm{n}(\bm{r}) + \rho_\mathrm{p}}{\rho_0(\bm{r})} - 1  \right)^2 \nonumber\\
&& + C_\mathrm{sat} \left( \rho_\mathrm{n}(\bm{r}) + \rho_\mathrm{p}(\bm{r})  \right) \left( \frac{\rho_\mathrm{n}(\bm{r}) + \rho_\mathrm{p}(\bm{r})}{\rho_0} - 1  \right),
\label{effpot}
\end{eqnarray}
where, $S_0$,  $K_0$,  $\epsilon_\mathrm{bind}$, and $C_\mathrm{sat}$ are adjustable parameters 
to satisfy the saturation properties of nuclear matter.

\subsection{Thermodynamic potential} \indent

\begin{wrapfigure}{l}{50mm}
%  \epsfxsize = \halftext
%  \centerline{ \epsfbox{edens_rho.eps}}
%\begin{figure}[h!]\begin{center}
%\includegraphics[width=8cm]{edens_rho.eps}
%\begin{center}
\includegraphics[width=45mm,height=70mm,keepaspectratio]{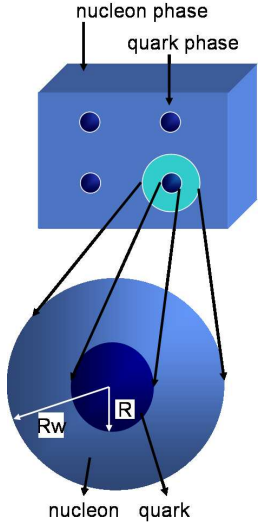}
\caption{ Wigner-Seitz approximation.}
\label{wsapp}
%\end{center}
%\end{figure}
\end{wrapfigure}

%We explain our framework to study the hadron-quark geometrical structured mixed phase. 
We consider a system which consists of the hadron and quark phases.
We divide the whole space into equivalent and charge-neutral Wigner-Seitz cells
with the size $R_W$  and the droplet size $R$  as illustrated in Fig.\ \ref{wsapp}.
Here we only consider the droplet phase.
Following DFT we begin with the thermodynamic potential in terms of
particle density profiles and chemical potentials, $\mu_i^\mathrm{Q}$ and $\mu_i^\mathrm{H}$,
\begin{equation}
\Omega_\mathrm{tot} = E(\rho_i(\bm{r}))-\sum_i\mu_i^\mathrm{Q} \int_0^R\!\! d\bm{r} \, \rho_i^\mathrm{Q}(\bm{r}) -\sum_i \mu_i^\mathrm{H} \int_R^{R_W}\!\! d\bm{r} \, \rho_i^\mathrm{H}(\bm{r}) .
\end{equation}
To take into account the Coulomb interaction, we write it in terms of the
$i$-particle
density profiles; $\rho_i(\bm{r})$ with $Q_i$ being the particle charge ($Q=-e
<0$ for the electron), 
\begin{equation}
V_{\mathrm{Coul}} (r) = -\sum_i \int d^3 r^{\prime} \frac{Q_i \rho_i(\bm{r}^{\prime})}{\left| \bm{r} - \bm{r}^{\prime} \right|}. \label{Coul}
\end{equation}
Applying the Laplacian $\nabla^2$ on Eq.\ (\ref{Coul}), we
can easily derive the Poisson equation $\nabla^2 V_\mathrm{Coul} = 4\pi
e^2 \sum_i Q_i \rho_i(\bm{r})$. 
The Coulomb interaction energy $E_V$ is then expressed as
\begin{equation}
E_V = \frac{1}{2} \sum_{i,j} \int d^3 r d^3 r^{\prime} \frac{Q_i \rho_i(\bm{r}) Q_j \rho_j(\bm{r}^{\prime})}{\left| \bm{r} - \bm{r}^{\prime} \right|}.
\end{equation}
Including the surface term, the total energy is expressed as
\begin{equation}
E(\rho_i(\bm{r})) = \int_0^R d\bm{r} \epsilon_\mathrm{Q} + \int_R^{R_W} d\bm{r} \epsilon_\mathrm{H} + \int_S dS \epsilon_S + E_V,
\label{totene}
\end{equation}
where $\epsilon_S$ means the surface-energy density, $S$ boundary area,
and $\epsilon_\mathrm{Q}$ and $\epsilon_\mathrm{H}$ the energy densities in the quark and hadron phases. 
Since we poorly know  
the details of the
hadron quark interface, we simply approximate the surface energy as $\displaystyle
\int_S dS \epsilon_S \equiv \sigma S $ by using the surface tension $\sigma$.
Each chemical potential is derived by the equation of motion
$\displaystyle \frac{\delta \Omega_\mathrm{tot}}{\delta \rho_i(\bm{r})} = 0$, 
which reads $\displaystyle \mu_i=\frac{\delta E(\rho_i(\bm{r}))}{\delta
\rho_i(\bm{r})}$, or
\begin{eqnarray}
\mu_i&=&\frac{\delta E_\mathrm{kin+str}}{\delta \rho_i(\bm{r})}-N_i V(\bm{r}), \hspace{10pt} N_i = \frac{Q_i}{e}, \label{chempot}\\
E_\mathrm{kin+str} &=& \int_0^R d\bm{r} \epsilon_\mathrm{Q} + \int_R^{R_W} d\bm{r} \epsilon_\mathrm{H} .
\end{eqnarray}

As we have seen in the previous section, each phase may have a finite net charge.
%%%%%%
%%%There are different charged phases, it is natural to occur the Coulomb interaction. 
%%%Many authors have included the Coulomb interaction,
%%%but its method is simply to add the Coulomb energy and the surface energy as
%%%the optimal value with respect to the volume fraction \cite{peth,alf2}.
%%%This method means ignoring the Coulomb screening effect.
In the previous studies of SMP, 
the Coulomb interaction between charged particles has been
treated rather simply: the Coulomb energy was
added to the total energy by using the volume fraction $f$ and constant 
densities derived
for two infinite matters \cite{peth,alf2}.
Thus the Coulomb screening effect and rearrangement of charge densities 
are completely discarded.
%%%%%%
One important point we would like to address here is, when there is the Coulomb interaction, 
the gauge variance of chemical potentials should be taken into account.
%%We can easily see the relation between chemical potentials and
%%the Coulomb potential under the gauge transformation.
%
%%Vary  
Differentiating 
the expression of chemical potential (\ref{chempot}) with respect to
the Coulomb potential $V(\bm{r})$, we get the relation \cite{vosk},
\begin{eqnarray}
A_{ij} \frac{\partial \rho_j}{\partial V} &=& N_i, \\
A_{ij} B_{jk} &=& \delta_{ik},
\end{eqnarray}
where matrices $A$ and $B$ are defined as 
\begin{eqnarray}
A_{ij} \equiv \frac{ \delta^2 E_{\mathrm{kin+str}}}{\delta \rho_i \delta \rho_j}, \ \ \ 
B_{ij} \equiv \frac{\partial \rho_i}{\partial \mu_j} .
\end{eqnarray}
 From these equations, the gauge invariance relation is derived,
\begin{equation}
\frac{\partial \rho_i}{\partial V} = N_j \frac{\partial \rho_j}{\partial \mu_i}.
\end{equation}
We can immediately see that the chemical potential is gauge variant from this
equation: when the Coulomb potential is
shifted by a constant value, $V(\bm{r}) \Longrightarrow V({\bm{r}}) - V_0$,
the chemical potential should also be shifted $\mu_i \Longrightarrow
\mu_i+N_i V_0$ to keep the gauge invariance. Note that the charge density 
should be observable and gauge
invariant; we shall see the constant shift of the Coulomb potential is compensated by the
redefinition of the chemical potential in the expression of the charge density. The Poisson equation is also
gauge invariant because it is not changed by the constant shift of the
Coulomb potential. 
Therefore, if the Poisson equation is not properly taken into account, the gauge
invariance is obviously violated.

Next, we present the explicit form of the thermodynamic potential for each particle. 
First, we consider the quark phase which consists of u, d and s quarks and 
electron.
According to the small current quark mass,  we treat u and d quarks as massless particles and only
s quark massive (150 MeV) in this phase.  Electron is also treated
as a massless particle. Next, we consider hadron phase which consists of 
proton,
neutron and electron. Nucleons are treated as non-relativistic particles. 
%Therefore 
The total thermodynamic potential becomes
\begin{equation}
\Omega_{\mathrm{tot}} = \Omega_{\mathrm{Q}} + \Omega_{\mathrm{H}}+ E_S. 
\end{equation}
Here $\Omega_{\rm Q}$ and $\Omega_{\rm H}$ are the thermodynamic
potentials in the quark and hadron phases,
%$\Omega_\mathrm{em}$ the contribution of electron and the
and $E_S$ the surface energy.

\subsubsection{
Quark phase (u, d, s quarks)}

The thermodynamic potentials 
%for quarks 
in the quark phase 
are given by Eqs.\ (\ref{kinene}), (\ref{fockene}), (\ref{totene}):
\begin{eqnarray}
\Omega_{\mathrm{Q}} &=& \Omega_{\mathrm{u}}+\Omega_{\mathrm{d}}+\Omega_{\mathrm{s}} + \Omega_\mathrm{em}^\mathrm{Q} + \int_0^R d\bm{r} B \hspace{10pt} \hbox{($B$ : bag constant)}. \label{omeq} \\
\Omega_{\mathrm{u}} &=& \int_0^R d^3r  \left[  \frac{3\pi^{\frac{2}{3}}}{4} \left( 1 + \frac{2 \alpha_s}{3 \pi} \right) \rho_{\mathrm{u}}^{\frac{4}{3}}(\bm{r}) - \mu_{\mathrm{u}} \rho_{\mathrm{u}}(\bm{r}) -\frac{2}{3}V_{\mathrm{Coul}}(\bm{r}) \rho_{\mathrm{u}}(\bm{r}) \right], \\
\Omega_{\mathrm{d}} &=& \int_0^R d^3r  \left[  \frac{3\pi^{\frac{2}{3}}}{4} \left( 1 + \frac{2 \alpha_s}{3 \pi} \right) \rho_{\mathrm{d}}^{\frac{4}{3}}(\bm{r}) - \mu_{\mathrm{d}} \rho_{\mathrm{d}}(\bm{r}) +\frac{1}{3}V_{\mathrm{Coul}}(\bm{r}) \rho_{\mathrm{d}}(\bm{r}) \right], \\
\Omega_{\mathrm{s}} &=&  \int_0^R d^3r  \left[  \epsilon_{\mathrm{s}}(\rho_{\mathrm{s}}(\bm{r})) - \mu_{\mathrm{s}} \rho_{\mathrm{s}}(\bm{r}) +\frac{1}{3}V_{\mathrm{Coul}}(\bm{r}) \rho_{\mathrm{s}}(\bm{r}) \right], \\
\Omega_\mathrm{em}^\mathrm{Q} &=& \int_0^{R} d^3r \left[ -\frac{1}{8\pi e^2} \left( \nabla V_{\mathrm{Coul}}(\bm{r}) \right)^2 + \frac{\left( 3 \pi^2 \rho_e(\bm{r}) \right)^\frac{4}{3}}{4\pi^2} - \mu_e \rho_e(\bm{r}) +  V_{\mathrm{Coul}}(\bm{r}) \rho_e(\bm{r})  \right].
\end{eqnarray}
Here, 
%u and d quarks are massless and only s quark is massive. 
$\Omega_\mathrm{em}^\mathrm{Q}$ summarizes the electron and
another Coulomb contributions.
The energy density of s quark
$\epsilon_{\mathrm{s}} (\rho_{\mathrm{s}}(\bm{r}))$ is a function of 
%%%the s quark density
$\rho_{\mathrm{s}}(\bm{r})$, which is explicitly expressed 
by using Eqs.\ (\ref{kinene}) and (\ref{fockene}) as 
\begin{eqnarray}
&&\epsilon_\mathrm{s}(\rho_\mathrm{s}(\bm{r})) 
= \left\{ \frac{3m_\mathrm{s}^4}{8\pi^2} \left[ \frac{p_\mathrm{Fs}(\bm{r})}{m_\mathrm{s}} 
  \sqrt{1 + \left(\frac{p_\mathrm{Fs}(\bm{r})}{m_\mathrm{s}} \right)^2}
  \left( 2 \left( \frac{p_\mathrm{Fs}(\bm{r})}{m_\mathrm{s}} \right)^2  + 1 \right)\right.\right. \nonumber\\
&&
  \left.\left.
  - \ln \left( \frac{p_\mathrm{Fs}(\bm{r})}{m_\mathrm{s}} 
    + \sqrt{ 1 + \left( \frac{p_\mathrm{Fs}(\bm{r})}{m_\mathrm{s}} \right)^2  }   \right)   \right] 
    - \frac{\alpha_s}{\pi^3} \left[ p_\mathrm{Fs}^4(\bm{r}) 
    - \frac{3}{2} m_\mathrm{s}^4 \left[ \frac{p_\mathrm{Fs}(\bm{r})}{m_\mathrm{s}} 
    \sqrt{ 1 + \left( \frac{p_\mathrm{Fs}(\bm{r})}{m_\mathrm{s}}\right)^2 } 
  \right.\right.\right.\nonumber\\
&& \left.\left.\left.
   - \ln \left( \frac{p_\mathrm{Fs}(\bm{r})}{m_\mathrm{s}} 
   +  \sqrt{ 1 + \left( \frac{p_\mathrm{Fs}(\bm{r})}{m_\mathrm{s}}\right)^2 }  \right)  \right]^2  \right] \right\} .
\end{eqnarray}

\subsubsection{
Hadron phase (non-relativistic nucleons)}

The thermodynamic potentials in the hadron phase read,
\begin{eqnarray}
\Omega_{\mathrm{H}} &=& \Omega_{\mathrm{n}} + \Omega_{\mathrm{p}} + \Omega_\mathrm{em}^\mathrm{H} + \int_R^{R_\mathrm{W}} d\bm{r} \epsilon_\mathrm{pot} (\bm{r}) \hspace{10pt} \hbox{here $\epsilon_\mathrm{pot}(\bm{r})$ is Eq.\ (\ref{effpot}).} \label{omeh}\\
%\Omega_{\mathrm{n}} &=&  2 \times  \int d^3r  \left[ \int_0^{p_{\mathrm{F}}  d^3p \sqrt{m_{\mathrm{n}}^2 + p^2 } - \left\{f\left( \rho_{\mathrm{p}}, \rho_{\mathrm{n}} \right)  \right\}  \rho_{\mathrm{n}} \right] \\
\Omega_{\mathrm{n}} &=& \int_R^{R_W} d^3r  \left[ \frac{3}{10m}\left( 3 \pi^2  \right)^\frac{2}{3} \rho^\frac{5}{3}_{\mathrm{n}}(\bm{r}) - \mu_{\mathrm{n}} \left( \rho_{\mathrm{p}}(\bm{r}), \rho_{\mathrm{n}}(\bm{r}) \right)   \rho_{\mathrm{n}}(\bm{r}) 
%\left( \rho_{\mathrm{p}}(\bm{r}), \rho_{\mathrm{n}}(\bm{r})  \right)  
\right], \\
%\Omega_{\mathrm{p}} &=& 2 \times \int d^3r  \left[  \int_0^{p_{\mathrm{F}}} d^3p \sqrt{m_{\mathrm{p}}^2 + p^2 } - \left\{ g\left( \rho_{\mathrm{p}}, \rho_{\mathrm{n}} \right)  - V_{\mathrm{Coul}} \right\}  \rho_{\mathrm{p}} \right] \\
\Omega_{\mathrm{p}} &=& \int_R^{R_W} d^3r  \left[   \frac{3}{10m}\left( 3 \pi^2  \right)^\frac{2}{3} \rho^\frac{5}{3}_{\mathrm{p}}(\bm{r}) - \mu_{\mathrm{p}}\left( \rho_{\mathrm{p}}(\bm{r}), \rho_{\mathrm{n}}(\bm{r}) \right)\rho_\mathrm{p}(\bm{r})  - V_{\mathrm{Coul}}(\bm{r}) \rho_{\mathrm{p}}(\bm{r}) \right], \\ 
\Omega_\mathrm{em}^\mathrm{H} &=& \int_R^{R_W} d^3r \left[ -\frac{1}{8\pi e^2} \left( \nabla V_{\mathrm{Coul}}(\bm{r}) \right)^2 + \frac{\left( 3 \pi^2 \rho_e(\bm{r}) \right)^\frac{4}{3}}{4\pi^2} - \mu_e \rho_e(\bm{r}) +  V_{\mathrm{Coul}}(\bm{r}) \rho_e(\bm{r})  \right].
\end{eqnarray}

Here $\Omega_\mathrm{em}^\mathrm{H}$ summarizes the electron and
another Coulomb contributions.

%Here, $g [\rho_{\mathrm{p}}, \rho_{\mathrm{n}}]$ is the function of $\rho_{\mathrm{p}}, \rho_{\mathrm{n}} $

%\subsubsection{
%Electron (in both Quark \& Hadron phases)}

%%Also there is electron in phase II. 
%Electrons exist in both phases.
%With the Coulomb potential,
%the electro-magnetic contribution to the thermodynamic potential becomes
%
%\begin{eqnarray}
%\Omega_{\mathrm{em}} &=& \int_0^{R_W} d^3r \left[ -\frac{1}{8\pi e^2} \left( \nabla V_{\mathrm{Coul}}(\bm{r}) \right)^2 + \frac{\left( 3 \pi^2 \rho_e(\bm{r}) \right)^\frac{4}{3}}{4\pi^2} - \mu_e \rho_e(\bm{r}) +  V_{\mathrm{Coul}}(\bm{r}) \rho_e(\bm{r})  \right] \nonumber \\
%   &=&   \int_0^{R_W} d^3r \left[ -\frac{1}{8\pi e^2} \left( \nabla V_{\mathrm{Coul}}(\bm{r}) \right)^2 - \frac{\left(V_{\mathrm{Coul}}(\bm{r}) - \mu_e \right)^4}{12\pi^2}  \right] .
%\end{eqnarray} 
%

\subsection{Equations of motion } \indent

We get the expression of chemical potentials and the Poisson
equation  from the equation of motion 
$\frac{\partial\Omega}{\partial\phi_i}=0$,
where $\phi_{i}=\rho_u(\bm{r}),
\rho_d(\bm{r}), \rho_s(\bm{r}), \rho_p(\bm{r}), \rho_n(\bm{r}), \rho_e(\bm{r}), V_{\mathrm{Coul}}(\bm{r})$. 
%%%These are the basic equations of motion in our present study.
%
The Poisson equation is explicitly written as
\begin{eqnarray}
\nabla^2 V_{\mathrm{Coul}}(\bm{r}) = 4 \pi e^2 \left[ \left(\frac{2}{3}\rho_u(\bm{r}) - \frac{1}{3}\rho_d(\bm{r}) - \frac{1}{3} \rho_s(\bm{r})  \right) \theta(R - r) + \rho_{\mathrm{p}}(\bm{r}) \theta(r - R) -  \rho_e(\bm{r})   \right].
\end{eqnarray}
The chemical potentials for quarks in the quark phase are derived as
\begin{eqnarray}
\mu_{\mathrm{u}} &=& \left( 1 + \frac{2 \alpha_s}{3 \pi}  \right) \pi^\frac{2}{3} \rho_{\mathrm{u}}^\frac{1}{3}(\bm{r}) - \frac{2}{3} V_{\mathrm{Coul}}(\bm{r}), \\
\mu_{\mathrm{d}} &=& \left( 1 + \frac{2 \alpha_s}{3 \pi}  \right) \pi^\frac{2}{3} \rho_{\mathrm{d}}^\frac{1}{3}(\bm{r}) + \frac{1}{3} V_{\mathrm{Coul}}(\bm{r}), \\
\mu_{\mathrm{s}} &=& \epsilon_{{\mathrm{Fs}}} + \frac{2 \alpha_s}{3 \pi} \left[ p_{\mathrm{Fs}}(\bm{r})- 3 \frac{m_{\mathrm{s}}^2}{\epsilon_{\mathrm{Fs}}} \ln \left( \frac{\epsilon_{\mathrm{Fs}}+p_{\mathrm{Fs}}(\bm{r})}{m_{\mathrm{s}}} \right)  \right]  + \frac{1}{3} V_{\mathrm{Coul}}(\bm{r}) .
\end{eqnarray}
 Here, we use
 $\epsilon_{\mathrm{Fs}} = \sqrt{m_{\mathrm{s}}^2+p_{\mathrm{Fs}}^2}$ and 
$p_{\mathrm{Fs}} = (\pi^2 \rho_{\mathrm{s}}(\bm{r}))^\frac{1}{3}$.
The nucleon chemical potentials in the hadron phase and the electron chemical potential are
\begin{eqnarray}
\mu_{\mathrm{n}} &=& \frac{p_{\mathrm{Fn}}^2}{2m} + \frac{2S_0 \left( \rho_n(\bm{r}) - \rho_{\mathrm{p}}(\bm{r}) \right)}{\rho_0} + \epsilon_{\mathrm{bind}} + \frac{K_0}{6} \left( \frac{\rho_{\mathrm{n}}(\bm{r}) + \rho_{\mathrm{p}}(\bm{r})}{\rho_0} - 1  \right)^2 \\
 &&+ \frac{K_0}{9} \left(  \frac{\rho_{\mathrm{n}}(\bm{r}) + \rho_{\mathrm{p}}(\bm{r})}{\rho_0}- 1  \right) + 2 C_{\mathrm{sat}}  \frac{\rho_{\mathrm{n}}(\bm{r}) + \rho_{\mathrm{p}}(\bm{r})}{\rho_0} - C_{\mathrm{sat}}, \\
\mu_{\mathrm{p}} &=& \mu_{\mathrm{n}} - \frac{p_{\mathrm{Fn}}^2}{2m} + \frac{p_{\mathrm{Fp}}^2}{2m} - \frac{4 S_0 \left( \rho_{\mathrm{B}}(\bm{r}) - 2 \rho_{\mathrm{p}}(\bm{r}) \right)^2}{\rho_0} - V_{\mathrm{Coul}}(\bm{r}), \\
\mu_e &=& \left( 3 \pi^2 \rho_e(\bm{r}) \right)^\frac{1}{3} + V_{\mathrm{Coul}}(\bm{r}).
\end{eqnarray}
Note that, e.g., the electron density  profile $\rho_\mathrm{e} (\bm{r})$ is expressed as
\begin{equation}
\rho_\mathrm{e} (\bm{r}) = \frac{\left( \mu_\mathrm{e} - V_\mathrm{Coul} (\bm{r}) \right)^3}{3\pi^2},
\end{equation}
in a gauge invariant fashion (cf.\ Eq.\ (\ref{mc2})).
%Here we remember the discussion of Eq.\ (\ref{mc}). That is, we can 
%recognize the electron density is depend on $\bm{r}$ because of
%introducing the Coulomb potential, while
%chemical potential is constant.
%
We impose the $\beta$ equilibrium in the quark and hadron phases and the chemical
equilibrium at the quark-hadron boundary as in Eqs.\ (\ref{chemequ})-(\ref{chemeqp}).
The pressure contribution from the surface tension is
derived as $P_\sigma = \sigma \frac{d S}{dV_\mathrm{Q}}$ with
$V_\mathrm{Q}$ being the volume of the quark droplet.
From Eqs.\ (\ref{omeq}) and (\ref{omeh}),  the pressure of each phase is expressed as
\begin{equation}
P^\mathrm{Q}=-\frac{\Omega_\mathrm{Q} }{V_\mathrm{Q}}, \hspace{5pt} P^\mathrm{H}=-\frac{\Omega_\mathrm{H} }{V_\mathrm{H}}, 
\end{equation}
with $V_\mathrm{H}$ being the volume of hadron phase.
Therefore the pressure balance condition becomes 
\begin{equation}
P^\mathrm{Q} = P^\mathrm{H} + P_\sigma,
\end{equation}
which gives the droplet size $R$. Finally the cell size $R_W$ is
determined by the minimum condition for $\Omega_\mathrm{tot}$.
Particle density profiles and $R$, $R_W$ are completely determined for
given $\mu_\mathrm{B}$.
Thus we can solve these coupled equations of motion consistently with GC. 
Note that
the Coulomb potential is included in a proper way and
$V_{\mathrm{Coul}}(\bm{r})$ appears in almost all chemical potentials.
The Coulomb potential is a
functional of the charged-particle density profiles and in turn
densities are functions of the Coulomb potential. As a result, the Poisson
equation becomes highly non-linear. 
Since it is difficult to solve analytically, we solve it numerically without any approximation,
whereas the linear approximation has been used in the previous work \cite{vosk}.

%\section{Maxwell construction}

\section{Numerical results} \indent

\begin{wrapfigure}{r}{80mm}
%  \epsfxsize = \halftext
%  \centerline{ \epsfbox{edens_rho.eps}}
%\begin{figure}[h!]\begin{center}
%\includegraphics[width=8cm]{edens_rho.eps}
%\begin{center}
\includegraphics[width=80mm]{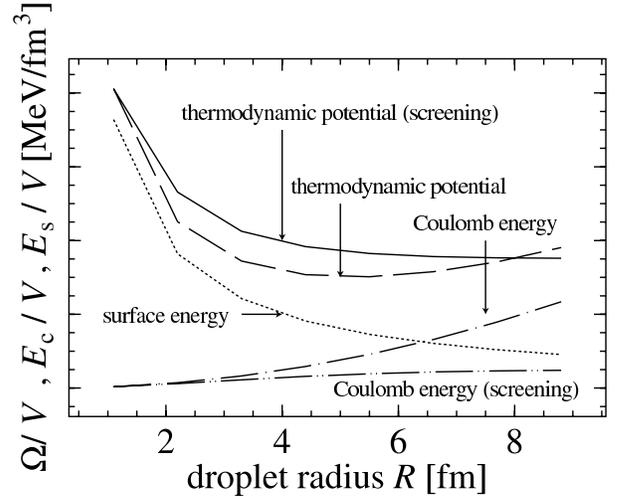}
\caption{ Radial dependence of the thermodynamic potential $\Omega$ for the droplet structure. 
Solid and dashed lines are $\Omega$ with and without screening, respectively.}
\label{insta}
%\end{center}
%\end{figure}
\end{wrapfigure}

We show a case of quark droplet embedded in hadron matter for the volume
fraction $f=(R/R_\mathrm{W})^3= 1/100$ and the surface tension $\sigma=60$ MeV/fm$^2$.
We can see the Coulomb screening effect
%\footnote{The Coulomb screening effect on quark droplet in vacuum is studied by Heiselberg \cite{heis}.}
in Fig.\ \ref{insta}. There is
always a minimum
in the thermodynamic potential with respect to the droplet radius $R$ without
the Coulomb screening, due to the balance between the Coulomb energy and
the surface energy. 
However, once the Coulomb screening 
%%%effect
is taken into account, the minimum 
disappears, which shows the mechanical instability caused by the Coulomb
screening effect. If the surface
tension is large enough, the geometrical structure (droplet)  becomes 
mechanically unstable by the Coulomb screening. 
%%%effect. 

\begin{figure}[htb]
\begin{minipage}[t]{80mm}
  \includegraphics[width=80mm]{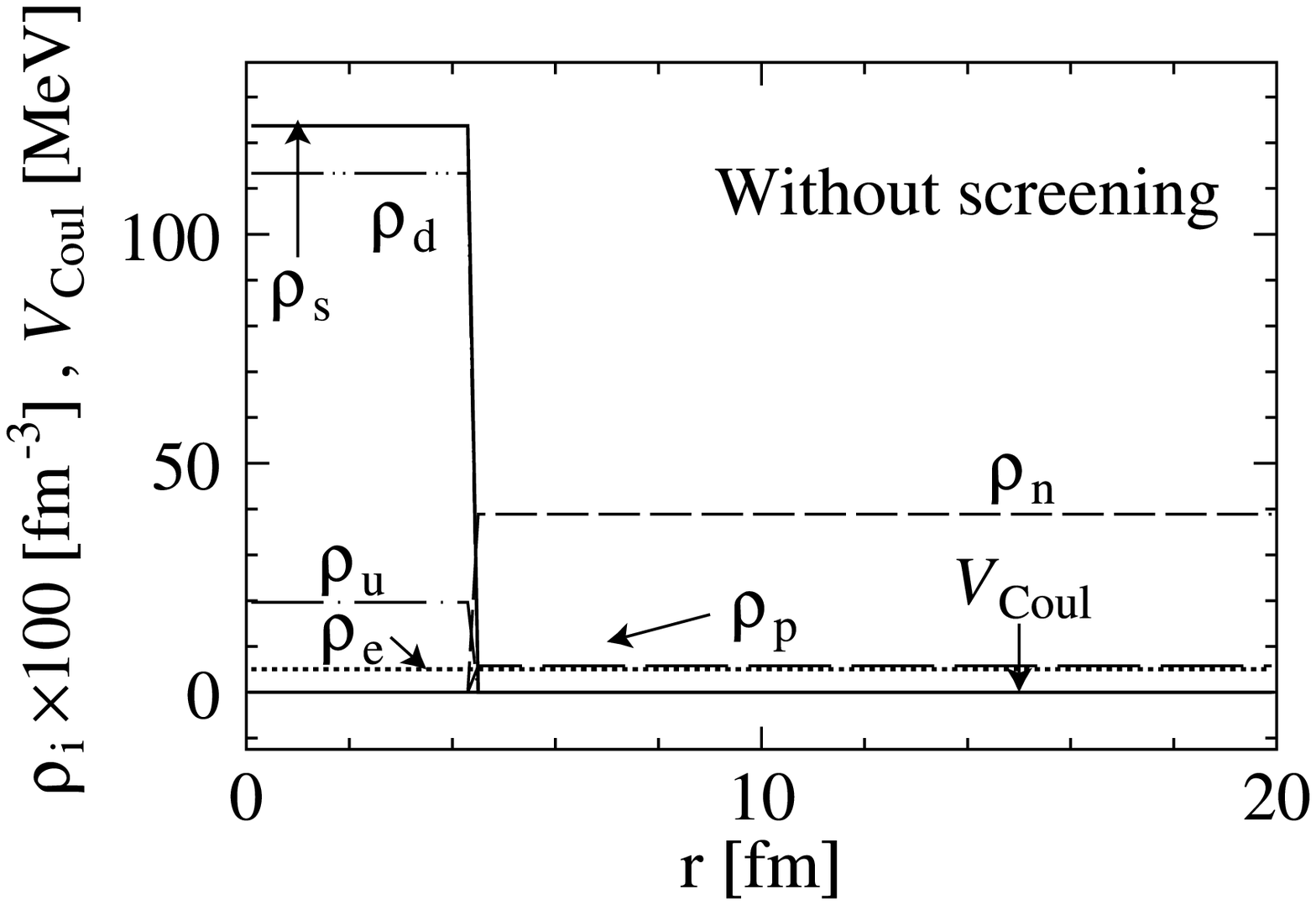}
%  \includegraphics[width=70mm]{sc.eps}
%\framebox[79mm]{\rule[-26mm]{0mm}{52mm}}
%\caption{Density profile of ``Without Screening effect'' in the case of
% $\mu_{\mathrm{B}}=1232$ MeV and $f=0.01$.  We
%defined volume fraction $f=\left(\frac{R}{R_{\mathrm{W}}}\right)^3$. Each density
% spreads constantly in each phase. Coulomb potential $V_{\mathrm{Coul}}$
% is constant (=0 MeV)}
%\label{fig:largenenough}
\end{minipage}
\hspace{8pt}
\begin{minipage}[t]{80mm}
  \includegraphics[width=80mm]{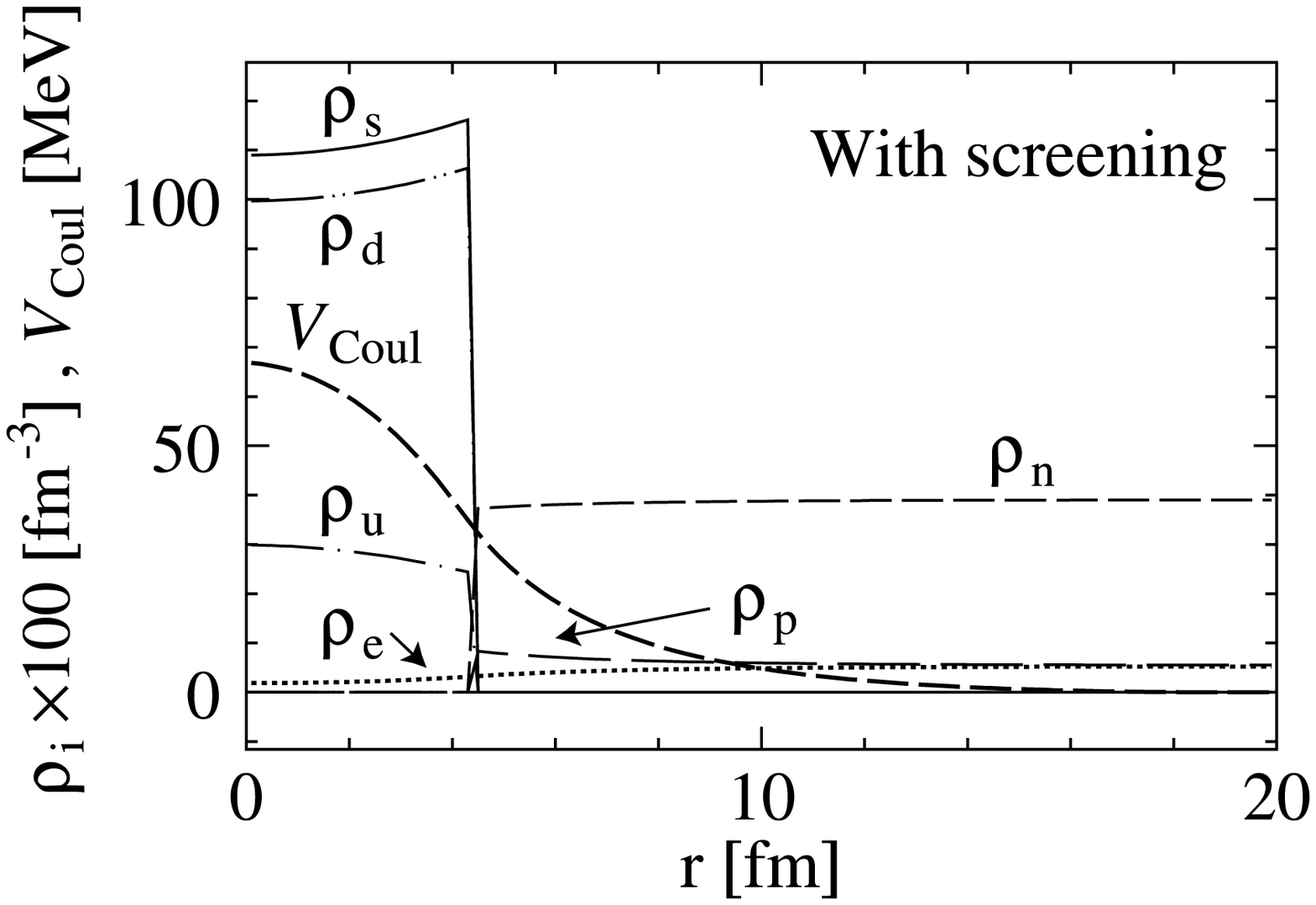}
%\caption{Density profile of ``With screening effect'' in the case of
% $\mu_{\mathrm{B}}=1232$ MeV and $f=0.01$. There is the
% density rearrangement by screening effect. We can see the feature
% that each phase tend to be charge neutral.}
%\label{fig:toosmall}
\end{minipage}
\caption{Density profiles $\rho_i$ without the Coulomb screening effect (left panel) and with
 the Coulomb screening effect (right panel). This is the case of
 $\mu_{\mathrm{B}}=1232$ MeV and the volume fraction
 $f=\left(R/R_{\mathrm{W}}\right)^3 = 0.01$. Each density uniformly
 spreads in each phase and $V_{\mathrm{Coul}}$ is constant ( =0 ) in the
 left panel, while there is rearrangement of the charge densities and
 $V_{\mathrm{Coul}}$ is spatially dependent in the right panel.}
\label{densprof}
\end{figure}

Next, we show the density profiles. We can see the
flat density profile of particles in the left panel of Fig.\ \ref{densprof},
where there is no Coulomb screening effect. 
In the right panel, on the other hand,  we can see the
rearrangement 
%%%effect 
of the charged particles 
%%%by 
as
the Coulomb screening effect.

Let us consider the features of these density profiles in detail. 
In the quark phase, the densities of the negatively charged  d, s quarks 
and electron are reduced, while
that of the positively charged u quarks are enhanced by the Coulomb 
screening effect. 
Remember that the quark phase has a negative charge and the hadron phase a positive charge in
the hadron-quark mixed phase. 

%By the Coulomb screening effect, the number of negatively charged particles are reduced 
%and that of positively particles are enhanced in the quark phase. 
For the hadron phase, we can see the opposite effect by the Coulomb screening:
the electron number increases and  the proton number is reduced. 
We see that the charged particles in one phase are also affected by the charge of another phase: 
negatively charged particles in the quark phase are attracted to the surface, 
%which cause the hadron phase to have positive charge, 
while positively charged particles (u quarks) are repelled from the surface. 
On the contrary, the positively charged particles in the hadron phase (protons) are attracted to the surface,
%which causes the quark phase to have negative charge,
 while electrons are repelled. 
Although the rearrangement effect is not enough to establish the local charge
neutrality, we can understand that the Coulomb screening effect is to
reduce the local charge in each phase.

 To draw the phase diagram
of the hadron-quark mixed phase, we have 
to minimize the thermodynamic potential with respect to the volume fraction, i.e., we have to change the droplet 
radius $R$ 
and Wigner-Seitz cell size $R_\mathrm{w}$.

\begin{figure}[htb]
\begin{minipage}[t]{80mm}
  \includegraphics[width=80mm]{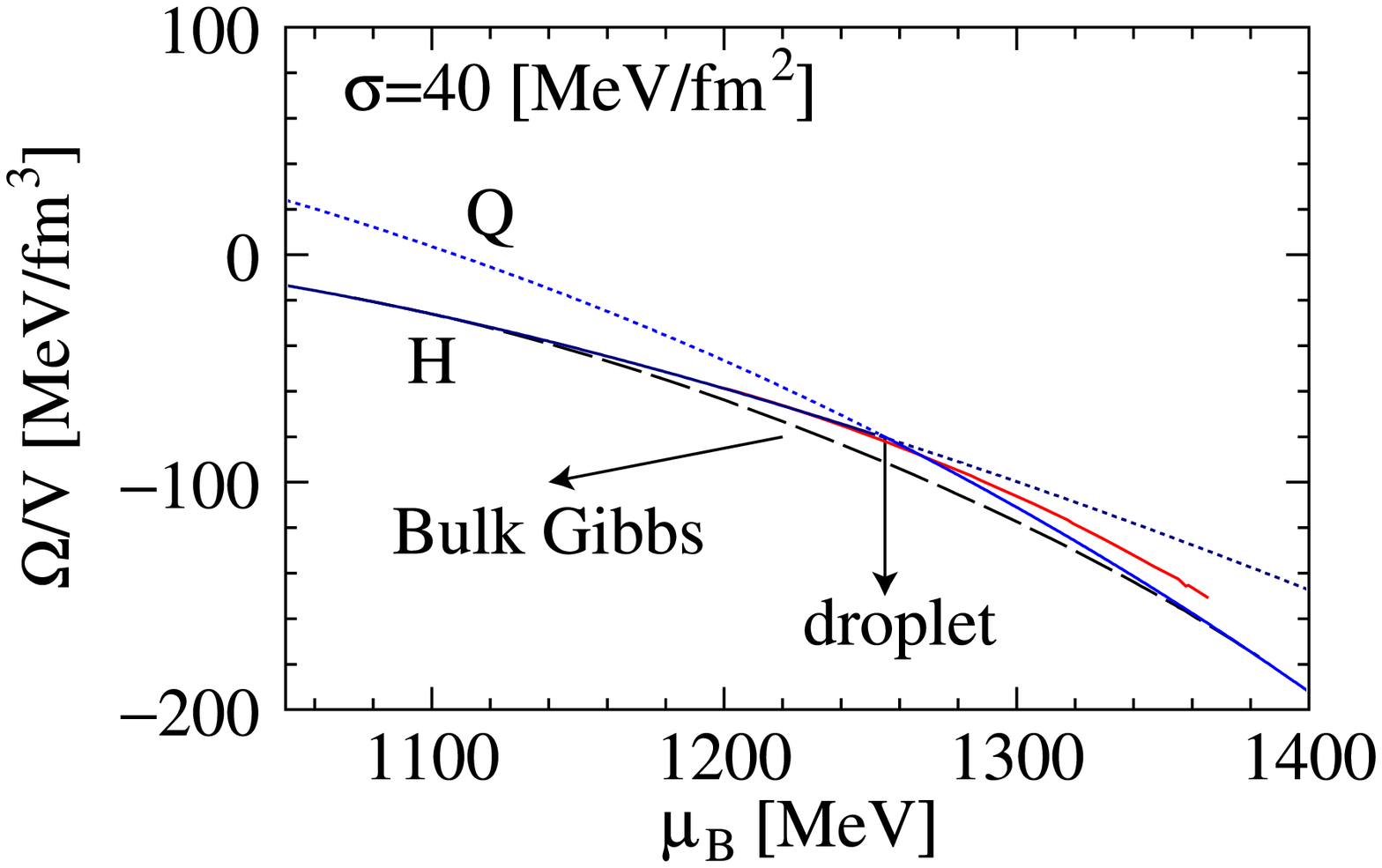}
\end{minipage}
\hspace{8pt}
\begin{minipage}[t]{80mm}
 \includegraphics[width=80mm]{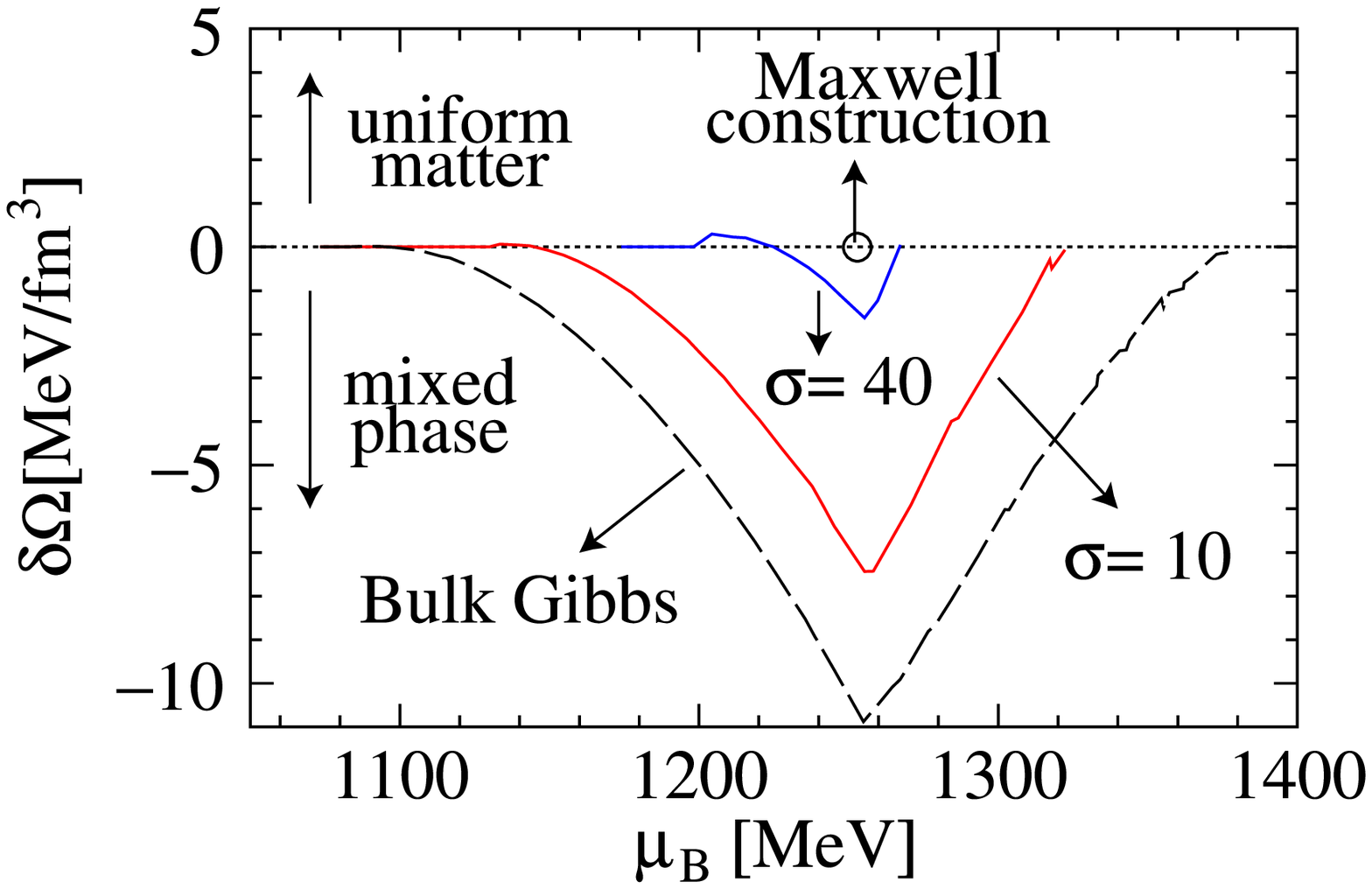}
\end{minipage}
\caption{ Thermodynamic potential $\Omega$ (left panel) and its difference from uniform hadron matter and
 quark matter $\delta \Omega$ (right panel). The dashed line ``Bulk
 Gibbs'', in the right panel, shows the
 difference between the bulk calculation and uniform matter. Solid lines
 are the cases for the droplet phase with $\sigma=10$
 and 40 MeV/fm$^2$.}
\label{omedef}
\end{figure}

We show the total thermodynamic potential for the droplet phase in
Fig.\ \ref{omedef}. Comparing with uniform hadron matter and quark
matter, the droplet phase takes
the smallest value of thermodynamic potential between a certain range of the baryon-number chemical potential. To
clarify the difference between uniform matter and the droplet phase,
we show the difference of thermodynamic potential between them
in the right panel of Fig.\ \ref{omedef}.  

We can see a large difference between our result ($\sigma=40$ MeV/fm$^2$) and
that given by the bulk calculation in Fig.\ \ref{omedef}. 

%%%The difference ``$\delta \Omega$'' shows the advantage or disadvantage of the mixed phase. 
If the difference $\delta \Omega$ is
negative, the mixed phase is an energetically favorable state. On the 
contrary, 
if $\delta \Omega$ is positive, the mixed phase is an unfavorable state.
In the right panel of Fig.\ \ref{omedef},
the point given by the Maxwell construction is specified by  a circle 
($\mu_{\mathrm{B}} \sim 1257$ MeV, $\delta \Omega =0$), 
where the following relations are maintained,
% because Maxwell construction is applied in achievement of GC;
\begin{equation}
 \mu_{\mathrm{B}}^{\mathrm{quark}} = \mu_{\mathrm{B}}^{\mathrm{hadron}} (\equiv \mu_{\mathrm{B}}),
 \hspace{5pt} P^{\mathrm{quark}}=P^{\mathrm{hadron}}, \hspace{5pt} T^{\mathrm{quark}} = T^{\mathrm{hadron}}.
\end{equation}
\noindent
Note that if we naively apply MC, we have to discard one of GC, $\mu_{\mathrm{e}}^{\mathrm{quark}} = \mu_{\mathrm{e}}^{\mathrm{hadron}}$ in the absence of the Coulomb potential.
The curve of ``Bulk Gibbs'' shows that the mixed phase is energetically
favorable in the wide $\mu_\mathrm{B}$ 
region,
as is already seen in Sec.\ 2.
The difference $\delta \Omega$ for 
``Bulk Gibbs'' is up to 10 MeV/fm$^3$, while for the
droplet phase ($\sigma=40$ MeV/fm$^2$), 
it is only less than 2 MeV/fm$^3$. This shows that
if we use the larger value of the surface tension, the mixed phase 
%%%%becomes to be unfavorable state. 
gets more unfavorable.
This feature is similar to that reported by Heiselberg \cite{peth} or
Alford \cite{alf2}. We can see the region of the energetically favorable mixed phase in
the right panel of Fig.\ \ref{omedef}: the region becomes narrower than
``Bulk Gibbs'', which means that the property of 
the mixed phase is closer to that given by MC. 
Thus we have seen that the mixed phase is very much affected by the Coulomb 
screening effect and the surface effect, by strictly keeping GC. For the 
larger value of the surface tension, MC is effectively useful in the description of the hadron-quark mixed phase. 
Note that we never violate the condition for the charge
chemical equilibrium
$\mu_\mathrm{e}^\mathrm{hadron}=\mu_\mathrm{e}^\mathrm{quark}$ 
to get these conclusions.
%: 
%GC (Eq.\ (\ref{gc})) have been strictly kept. 
Remember that the meaning of the condition 
in the presence of the Coulomb potential 
is different 
from that in the absence of the Coulomb potential. 
%we pointed out it can be possible $\mu_\mathrm{e}^\mathrm{hadron}
%=\mu_\mathrm{e}^\mathrm{quark}$ may be fulfilled in MC (Eq.\ (\ref{mc})) 
%by introducing the Coulomb
%potential in a proper way. This suggest that
  
%This feature mainly comes from surface energy. 

%The references must be written with the following style
%\cite{1}-\cite{4}

%\centerline{\bf Figure sample}

%\begin{figure}[h!]\begin{center}
%\includegraphics[width=8cm]{nic8.eps}
%\caption{ Figure Caption}
%\label{fig1}\end{center}
%\end{figure}

\section{Summary and concluding remarks} \indent
 
In this note we have first seen how the bulk calculation is performed for the 
hadron-quark mixed phase by applying the Gibbs conditions to the system 
consisting of two infinite matters. It gives a wide density region for the 
mixed phase. Based on this result, some authors also suggested the  
structured mixed phase with various geometrical structures, by including 
the finite-size effects, the surface and Coulomb energies \cite{peth,alf2}. 

We have examined the finite-size effects in the mixed phase by numerically 
solving the equations 
of motion for the particle densities and the Poisson equation for the Coulomb 
potential. Our framework is based on the idea of the density functional 
theory, which, we believe, is one of the best theories to treat the 
structured mixed phases. We have demonstrated, by taking the droplet phase 
as an example, that the Coulomb screening effect and 
 rearrangement of the charge densities play
an important role for the 
mechanical instability as well as the energy of the mixed phase. As a result 
we have seen that the region of the mixed phase is highly restricted by the 
Coulomb screening effect as well as the surface energy. We have also seen 
that EOS gives the similar behavior to that given by the Maxwell 
construction, whereas the Maxwell construction is apparently incorrect 
in a system with more than one chemical potential: the Maxwell construction 
is effectively useful in the description of the mixed phase, even in this 
case.    
%We have demonstrated the simple model calculation: ``Bulk Gibbs'' and
%seen mixed phase appears in a wide density region. Although many authors
%have used bulk calculation including the surface and coulomb effect, 
%we studied them in a proper way. 
%We have carefully taken into account the Coulomb
%potential and have seen the rearrangement of charged particles 
%caused by the screening effect. We have seen that SMP, here we show the 
%droplet phase, is
%mechanically unstable if the surface tension is large. As a result mixed
%phase region becomes narrow and it could be said that the screening 
%effect makes the system similar to
%the picture derived from MC. 
%Although we know that MC is apparently incorrect in  a system with
% more than one chemical potential, we can say that MC is
%effectively useful
%in description of the mixed phase.
% where MC is meaningless at a first glance. 
As another case of more than one chemical potential, kaon condensation  
has been also studied \cite{maru} and the result is similar to 
the present case. 

We have included the surface tension, but its definite value is not clear
 and many authors treated it as a free parameter. 
There are also many
estimations for the surface tension at the hadron-quark
interface in lattice QCD \cite{kaja,huan}, in shell-model
calculations \cite{mad1,mad2,berg} and in model
calculations based on the Dual-Ginzburg Landau theory \cite{mond}.
%There are also many
%estimations for the surface tension at the hadron-quark
%interface: a rough estimation suggests 300 MeV/fm$^2$
%from the difference of energy density between hadron-quark phases. However,
%surface tension should also depend on many other contributions, for
%example the structure of mixed phase. Many calculation have been done and
%now it is estimated as 10-60 MeV/fm$^2$ in lattice QCD \cite{kaja,huan}, 30-40 MeV/fm$^2$ in shell-model
%calculations \cite{mad1,mad2,berg}, and 25-50 MeV/fm$^2$ in model
%calculations based on the Dual-Ginzburg Landau theory \cite{mond}.
If we have the realistic value of the surface tension, we can reasonably bring out SMP in
the hadron-quark phase transition. 
As these system
corresponds to neutron star matter, we have seen that the mixed phase
should be narrow by the finite-size effects. Our result would restrict
the allowed SMP region in neutron stars which is suggested by Glendenning
\cite {gle2,gle3}. It could be said that
they should change the
property of neutron stars especially in equation of state \cite{end2}.

We have assumed in relation to phenomenological implications here that
temperature is zero. It would be  much interesting to include the finite-temperature
effect. Then it is possible to draw the phase diagram in the
$\mu_\mathrm{B}-T$ plane and we can study the properties of the
deconfinement phase transition.
 In this study we have used a simple model for quark matter to
figure out the finite-size effects in the structured mixed phase. However,
it has been suggested that the color superconductivity is a ground state of quark
matter \cite{alf1,alf2}. To get more realistic picture of
the hadron-quark phase
transition, we will need to
take into account color superconductivity. In the recent studies the
mixed phase has been also studied \cite{shov,redd}.

\section*{Acknowledgments} \indent

We thank to D.~N.~Voskresensky and T.~Tanigawa for their fruitful discussion.
T.E.\ and T.T.\ acknowledge the support and kind hospitality of 
the Joint Institute
for Nuclear Research (JINR) when we stayed in Dubna.

%\subsubsection*{Appendix}
%Here is the Appendix.

\end{document}